\documentclass[reprint,twocolumn]{revtex4}%
\usepackage{amsfonts}
\usepackage{amsmath}
\usepackage{amssymb}
\usepackage{charter}
\usepackage{graphicx}%
\providecommand{\U}[1]{\protect \rule{.1in}{.1in}}
\begin{document}
\title{Laguerre polynomial excited coherent states generated by multiphoton
catalysis: Nonclassicality and Decoherence}
\author{{\small Li-Yun Hu}$^{1,2\ast}${\small \thanks{Corresponding author. Email:
hlyun2008@126.com}, Jia-Ni Wu}$^{1}${\small , Zeyang Liao}$^{2}${\small , and
M. S. Zubairy}$^{2}$}
\affiliation{$^{1}${\small Center for Quantum Science and Technology, Jiangxi Normal
University, Nanchang 330022, China}}
\affiliation{$^{2}${\small Institute for Quantum Science and Engineering (IQSE) and
Department of Physics and Astronomy, Texas A\&M University, College Station,
TX 77843, USA}}

\begin{abstract}
We theoretically introduce a new kind of non-Gaussian state-----Laguerre
polynomial excited coherent states by using the multiphoton catalysis which
actually can be considered as a block comprising photon number operator. It is
found that the normalized factor is related to the two-variable Hermite
polynomials. We then investigate the nonclassical properties in terms of
Mandel's Q parameter, quadrature squeezing, second correlation, and the
negativity of Wigner function (WF). It is shown that all these properties are
related to the amplitude of coherent state, catalysis number and unbalanced
beam splitter (BS). In particular, the maximum degree of squeezing can be
enhanced as catalysis number and keeps a constant for single-photon catalysis.
In addition, we examine the effect of decoherence by Wigner function, which
show that the negative region, characteristic time of decoherence and
structure of WF are affected by catalysis number and unbalanced BS. Our work
provides a general analysis about how to prepare theoretically polynomials
quantum states.

{\small Keywords: nonclassical property; Laguerre polynomials excitation;
squeezing; Wigner function}

{\small PACS: 03.65 -a. 42.50.Dv}

\end{abstract}
\maketitle

\section{Introduction}

Nonclassical state has an important role in understanding deeply some
fundamental problems in the field of quantum mechanics. In order to realize
this purpose, many experimental and theoretical protocols have been proposed
to generate and manipulate such nonclassical quantum states
\cite{1,2,2a,3,4,4a,5,6,6a,7}. In these protocols, the photon addition
$a^{\dag}$, as a non-Gaussian operation, can create a nonclassical state from
any classical state \cite{2,7}. In addition, practically realizable
non-Gaussian operations including the photon subtraction $a$ or addition
$a^{\dag}$ or the superposition of both were used to improve the
nonclassicality, the degree of entanglement, the fidelity of continuous
variable teleportation, loophole-free tests of Bell's inequality, and quantum
computing, as well as the performance of quantum-key-distribution
\cite{3,8,9,10,11,12,13,14}. For example, the quantum commutation rules have
been probed experimentally by using addition and subtraction of single photons
to/from a light field \cite{4,4a}.

In addition, the multi-photon process has experimentally an theoretically
attracted much attention \cite{7,15,16,17,18,19,20,21}. For example, the
multi-photon excited coherent state has been introduced by Agarwl, and the
corresponding nonclassical properties and experimental preparation are
discussed by using parametric down-conversion and homodyne tomography
technology \cite{7,22,23,24,25,26,27}. Single photon addition is used
theoretically to improve the performance of quantum-key-distribution
\cite{13}. Recently, multiple-photon subtraction and addition have been used
to enhance the degree of entanglement for two-mode squeezed vacuum (TMSV) and
the fidelity of teleportation with continuous \cite{19,20,28,29}. It is shown
that the highest entanglement, the fidelity and some squeezing properties can
be improved for the TMSV with symmetric multi-photon subtraction operations.
Both two non-Gaussian operations can actually be realized by using the linear
optical elements such as beam splitter and conditional measurement on
ancillary outcome, which is probabilistic but more feasible in the laboratory
compared with nonlinear process.

On the other hand, it is interesting to notice that Hermite polynomial states
can be considered as the minimum uncertain states \cite{30,31}, on which are
focused by some researchers \cite{14,32,33}. For instance, a generalized
Hermite polynomial's operation has been theoretically introduced and operated
on single-mode squeezed vacuum and coherent state \cite{32,33}, such as
$H_{n}\left(  Q\right)  \left \vert \alpha \right \rangle $ and $H_{n}\left(  \mu
a+\nu a^{\dagger}\right)  S\left(  r\right)  \left \vert 0\right \rangle $ where
$Q=(a+a^{\dagger})/\sqrt{2}$ is the coordinate operator and $\left \vert
\alpha \right \rangle $ is the Glauber coherent state, and $S\left(  r\right)
\left \vert 0\right \rangle $ is the single-mode squeezed vacuum. It is found
that all these nonclassicalities can be enhanced by Hermite polynomial
operation and two adjustable parameters \cite{33}. However, there is no scheme
proposed to generate such these polynomial states. The implementation of such
non-Gaussian operations is still a very challenging task \cite{34}.

In order to prepare the non-Gaussian states, excepting for photon addition and
subtraction or both, the quantum catalysis is also a feasible strategy to
generate nonclassical quantum states \cite{35,17}. The analogy to catalysis is
to perform a measurement with the same number of photons as ancillary mode on
one output, which can generate an effective nonlinearity. In this paper, we
shall introduce a new kind of non-Gaussian quantum states-----Laguerre
polynomial excited coherent states (LPECSs), which can be produced by using
beam splitter and a special conditional measurement (multi-photon catalysis)
on one of two outports. Then we investigate the nonclassical properties
according to the Mandel's Q parameter and second-order correlation function,
photon-number distribution, squeezing property as well as the Wigner function.
Particularly, we also discuss the decoherence effect of thermal channel on the
LPECSs by deriving analytically the Wigner function. There is no report about
this non-Gaussian state generated by multiphoton catalysis before, including
the effect of decoherence on the nonclassicality.

This work is arranged as follows. In Sec. II, we propose the protocol for
generating such a kind of non-Gaussian state by using the heralded
interference and conditional measurement. In Sec. III, we derive the
normalization factor, which is important for further discussing the
statistical properties of the state. It is shown that the factor is related to
the two-variable Hermite polynomials. In Sec. IV, we present the statistical
properties of the state, such as photon-number distribution, squeezing, etc.
Secs. V and VI are devoted to investigating the nonclassicality in terms of
the negativity of Wigner function without and with the effect of decoherence
of the thermal channel, respectively. Our conclusions and discussions are
presented in the last section.

\section{The generation of the LPECSs}

The scheme for generating an optical state $\left \vert \Psi \right \rangle
_{out}$ by the heralded interference and conditional measuring $m$ photons is
shown in Fig.1. In Fig.1, an arbitrary input pure state $\left \vert
\varphi \right \rangle _{in}$ and an $m-$photon Fock state $\left \vert
m\right \rangle _{a}$ are sent on an asymmetrical beam splitter (BS), and a
number measurement is performed on one of the two outports.
\begin{figure}[ptb]
\label{Fig1} \centering \includegraphics[width=6cm]{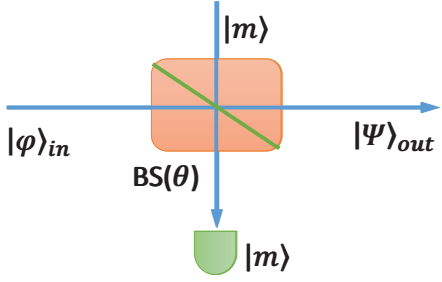}\caption{{}(Color
online) An optical state $\left \vert \Psi \right \rangle _{out}$ is generated by
the interference between an arbitrary input pure state $\left \vert
\varphi \right \rangle _{in}$ and an $m-$photon Fock state (at a beam splitter
of reflectivity $\left \vert r\right \vert ^{2}=1-\left \vert t\right \vert ^{2}%
$), conditional measuring $m$ photons at one output.}%
\end{figure}

If we have a conditional measurement with $m$ photons at one output port (see
Fig.1), then the conditioned state at the other output is given by%
\begin{equation}
\left \vert \Psi \right \rangle _{out}=N_{m}\left.  _{a}\left \langle m\right \vert
B\left(  \theta \right)  \left \vert m\right \rangle _{a}\right.  \left \vert
\varphi \right \rangle _{in},\label{1}%
\end{equation}
where $N_{m}$ is the normalization factor, and $B\left(  \theta \right)
=\exp \left \{  \theta(a^{\dagger}b-ab^{\dagger})\right \}  $ is the BS operator,
and $r=\sin \theta$, $t=\cos \theta$. BS operator is actually an entangling
operator \cite{36a}.When $\theta=\pi/4$, $B\left(  \pi/4\right)  $ is the
symmetrical BS. In order to further obtain the expression in Eq.(\ref{1}), we
first derive the matrix element $_{a}\left \langle m\right \vert B\left(
\theta \right)  \left \vert m\right \rangle _{a}$. Using the normal ordering form
of $B\left(  \theta \right)  $ \cite{36}:%
\begin{align}
B\left(  \theta \right)   &  =\colon \exp \left \{  \left(  \cos \theta-1\right)
\left(  a^{\dag}a+b^{\dag}b\right)  \right.  \nonumber \\
&  +\left.  \left(  a^{\dagger}b-ab^{\dagger}\right)  \sin \theta \right \}
\colon,\label{2}%
\end{align}
and the coherent state representation of Fock state, i.e.,
\begin{equation}
\left \vert m\right \rangle =\frac{1}{\sqrt{m!}}\frac{\partial^{m}}%
{\partial \alpha^{m}}\left.  \left \Vert \alpha \right \rangle \right \vert
_{\alpha=0},\left \Vert \alpha \right \rangle =\exp \left(  \alpha a^{\dagger
}\right)  \left \vert 0\right \rangle ,\label{3}%
\end{equation}
we can derive
\begin{align}
\hat{B}_{m} &  \equiv \left.  _{a}\left \langle m\right \vert B\left(
\theta \right)  \left \vert m\right \rangle _{a}\right.  \nonumber \\
&  =\frac{\left(  -\cos \theta \right)  ^{m}}{m!}\colon H_{m,m}\left(  b^{\dag
}\tan \theta,b\tan \theta \right)  \colon e^{b^{\dag}b\ln \cos \theta}\nonumber \\
&  =\cos^{m}\theta \colon L_{m}\left(  b^{\dag}b\tan^{2}\theta \right)  \colon
e^{b^{\dag}b\ln \cos \theta},\label{4}%
\end{align}
where $L_{m}\left(  .\right)  $ is the Laguerre polynomials, $H_{m,m}\left(
x,y\right)  $ is the two-variable Hermite polynomials, and we have used the
operator identity $\colon \exp \{ \left(  e^{\lambda}-1\right)  b^{\dag}%
b\} \colon=e^{\lambda b^{\dag}b}$ and $e^{\lambda b^{\dag}b}be^{-\lambda
b^{\dag}b}=e^{-\lambda}b$ and the relation $\left(  -1\right)  ^{m}%
/m!H_{m,m}\left(  x,y\right)  =L_{m}\left(  xy\right)  $. Thus, for any input
state, the output state can be expressed as $\left \vert \Psi \right \rangle
_{out}\rightarrow \hat{B}_{m}\left \vert \varphi \right \rangle _{in}$. It will be
convenient to further discuss some properties of the output states by using
Eq.(\ref{4}). From Eq.(\ref{4}), we can see that the process, accompanying
with $m$-photon Fock state input and $m$-photon measured, can be seen as a
kind of Laguerre polynomials operation of number operator within normal
ordering form.

When the input state $\left \vert \varphi \right \rangle _{in}$ is the coherent
state $\left \vert z\right \rangle $, then the output state is given by
\begin{align}
\left \vert \Psi \right \rangle _{out} &  =N_{m}\cos^{m}\theta \colon L_{m}\left(
b^{\dag}b\tan^{2}\theta \right)  \colon e^{b^{\dag}b\ln \cos \theta}\left \vert
z\right \rangle \nonumber \\
&  =N_{m}e^{-\frac{1}{2}\left \vert z\right \vert ^{2}\sin^{2}\theta}\cos
^{m}\theta L_{m}\left(  \mu b^{\dag}\right)  \left \vert z\cos \theta
\right \rangle \nonumber \\
&  \equiv \bar{N}_{m}L_{m}\left(  \mu b^{\dag}\right)  \left \vert z\cos
\theta \right \rangle ,\label{5}%
\end{align}
where we have set $\mu=z\cos \theta \tan^{2}\theta,$ $\bar{N}_{m}=N_{m}\cos
^{m}\theta \exp[-\frac{1}{2}\left \vert z\right \vert ^{2}\sin^{2}\theta]$ and we
have used the formula
\begin{equation}
g^{b^{\dag}b}\left \vert \alpha \right \rangle =\exp \left[  \frac{1}{2}\left(
g^{2}-1\right)  \left \vert \alpha \right \vert ^{2}\right]  \left \vert
g\alpha \right \rangle .\label{6}%
\end{equation}
It is clear that Eq.(\ref{5}) is just the Laguerre polynomials excited
coherent state (LPECSs) generated by the condition measurement. Let us note
that the scaling $\cos \theta$ of the coherent state $z$ can be understood as a
loss process. This character is a result of the process itself. For the case
of $\theta=0$ corresponding to the perfect transmission ($t=1,r=0$), we see
$\left \vert \Psi \right \rangle _{out}\rightarrow \left \vert z\right \rangle $, as
expected. When $m=0$, $1$, the output states become $\left \vert \Psi
\right \rangle _{out}=\left \vert z\cos \theta \right \rangle $, and $\left \vert
\Psi \right \rangle _{out}=\bar{N}_{1}\left(  1-\mu b^{\dag}\right)  \left \vert
z\cos \theta \right \rangle $, respectively. The former is still a coherent state
with a smaller amplitude $z\cos \theta$ comparing with that of the initial
input state, which means that even when for instance $m=0$ (without photon
detected) the average number of photon at the output is $\left \vert
z\right \vert ^{2}\cos^{2}\theta,$ not $\left \vert z\right \vert ^{2},$ i.e.,
the average numbers of photon are not conservation for the input-output state
at the process of quantum catalysis; And the latter corresponds to a
superposition of coherent state and excited coherent state.

\section{Normalization of the LPECSs}

Next, we derive the normalization of the LPECSs, which is important for
discussing the statistical properties of quantum states. Using the normalized
condition $1=\left.  _{out}\left \langle \Psi \right \vert \left.  \Psi
\right \rangle _{out}\right.  $ and the completeness relation $\int d^{2}%
\alpha \left \vert \alpha \right \rangle \left \langle \alpha \right \vert /\pi=1$ of
coherent state, as well as
\begin{equation}
\left \langle z\cos \theta \right \vert \left.  \alpha \right \rangle =\exp \left \{
-\frac{\left \vert z\right \vert ^{2}}{2}\cos^{2}\theta-\frac{1}{2}\left \vert
\alpha \right \vert ^{2}+z^{\ast}\alpha \cos \theta \right \}  , \label{7}%
\end{equation}
we can derive%
\begin{align}
\bar{N}_{m}^{-2}  &  =\left \langle z\cos \theta \right \vert L_{m}\left(
\mu^{\ast}b\right)  L_{m}\left(  \mu b^{\dag}\right)  \left \vert z\cos
\theta \right \rangle \nonumber \\
&  =\int \frac{d^{2}\alpha}{\pi}\left \vert L_{m}\left(  \mu^{\ast}%
\alpha \right)  \right \vert ^{2}\left \vert \left \langle z\cos \theta \right \vert
\left.  \alpha \right \rangle \right \vert ^{2}\nonumber \\
&  =\int \frac{d^{2}\alpha}{\pi}\left \vert L_{m}\left(  \mu^{\ast}%
\alpha \right)  \right \vert ^{2}e^{-\left \vert z\right \vert ^{2}\cos^{2}%
\theta-\left \vert \alpha \right \vert ^{2}+\left(  z^{\ast}\alpha+z\alpha^{\ast
}\right)  \cos \theta}. \label{8}%
\end{align}
Using the sum representation of Laguerre polynomial
\begin{equation}
L_{m}\left(  x\right)  =\sum_{l=0}^{m}\binom{m}{l}\frac{(-1)^{l}}{l!}x^{l},
\label{9}%
\end{equation}
we can rewrite Eq.(\ref{8}) as the following form%
\begin{align}
\bar{N}_{m}^{-2}  &  =\sum_{l,k=0}^{m}\binom{m}{l}\binom{m}{k}\frac
{(-1)^{l+k}}{l!k!}\mu^{k}\mu^{\ast l}\nonumber \\
&  \times e^{-\left \vert z\right \vert ^{2}\cos^{2}\theta}\int \frac{d^{2}%
\alpha}{\pi}\alpha^{\ast k}\alpha^{l}e^{-\left \vert \alpha \right \vert
^{2}+\left(  z^{\ast}\alpha+z\alpha^{\ast}\right)  \cos \theta}\nonumber \\
&  =\sum_{l,k=0}^{m}\binom{m}{l}\binom{m}{k}\frac{(-1)^{k}\mu^{k}\mu^{\ast l}%
}{l!k!}H_{k,l}\left(  z^{\ast}\cos \theta,-z\cos \theta \right)  , \label{10}%
\end{align}
where we have used the following integration formula%
\begin{equation}
H_{m,n}\left(  \xi,\eta \right)  =(-1)^{n}e^{\xi \eta}\int \frac{d^{2}z}{\pi
}z^{n}z^{\ast m}e^{-\left \vert z\right \vert ^{2}+\xi z-\eta z^{\ast}}.
\label{11}%
\end{equation}
Eq.(\ref{10}) is the analytical expression of normalization factor for the
output state $\left \vert \Psi \right \rangle _{out}$, which is related to the
two-variable Hermite polynomials. $\bar{N}_{m}^{-2}$ is a real number which
can be seen directly from Eq.(\ref{8}). In particular, when $m=1$
corresponding to the single-photon catalysis, we have $\bar{N}_{1}%
^{-2}=(1-\left \vert z\right \vert ^{2}\sin^{2}\theta)^{2}+\left \vert
z\right \vert ^{2}\cos^{2}\theta \tan^{4}\theta$ which is in accordance with
\cite{17}.

In a similar way to deriving Eq.(\ref{10}), we can calculate the matrix
element $\left \langle b^{q}b^{\dag p}\right \rangle $\ as
\begin{align}
\left \langle b^{q}b^{\dag p}\right \rangle  &  =\sum_{l,k=0}^{m}\binom{m}%
{l}\binom{m}{k}\frac{(-1)^{q+k}}{l!k!}\mu^{k}\mu^{\ast l}\nonumber \\
&  \times \bar{N}_{m}^{2}H_{k+p,l+q}\left(  z^{\ast}\cos \theta,-z\cos
\theta \right)  , \label{12}%
\end{align}
which will be often used in the next calculation for discussing the
nonclassical properties of the LPECSs.

\section{Nonclassical properties of the LPECSs}

In this section, we shall discuss the nonclassical properties of the LPECSs by
using Mandel's $\mathfrak{Q}$ parameter and second-order correlation function,
photon-number distribution, as well as squeezing property.

\subsection{Mandel's $\mathfrak{Q}$ parameter}

First, let us examine the sub-Possion statistical property using the Mandel
$\mathfrak{Q}$-parameter \cite{37}, whose definition can be given by%
\begin{align}
\mathfrak{Q}  &  =\frac{\left \langle \left(  b^{\dag}b\right)  ^{2}%
\right \rangle -\left \langle b^{\dag}b\right \rangle ^{2}}{\left \langle b^{\dag
}b\right \rangle }-1\nonumber \\
&  =\frac{\left \langle b^{2}b^{\dag2}\right \rangle -\left \langle bb^{\dag
}\right \rangle ^{2}-2\left \langle bb^{\dag}\right \rangle +1}{\left \langle
bb^{\dag}\right \rangle -1}. \label{13}%
\end{align}
The quantum state shall satisfy the sub-Poissonian statistics when the
condition $\mathfrak{Q}<0$ is achieved. The super-Poissonian, Poissonian
statistics correspond to $\mathfrak{Q}>0$ and $\mathfrak{Q}=0$, respectively.
For simplicity, here we have converted the expression of $\mathfrak{Q}$ to the
anti-normally ordering form. Using Eq.(\ref{12}), we can get the analytical
expression of $\mathfrak{Q}$ but do not give them here due to its long and cumbersome.

\begin{figure}[ptb]
\label{Fig2} \centering \includegraphics[width=6cm]{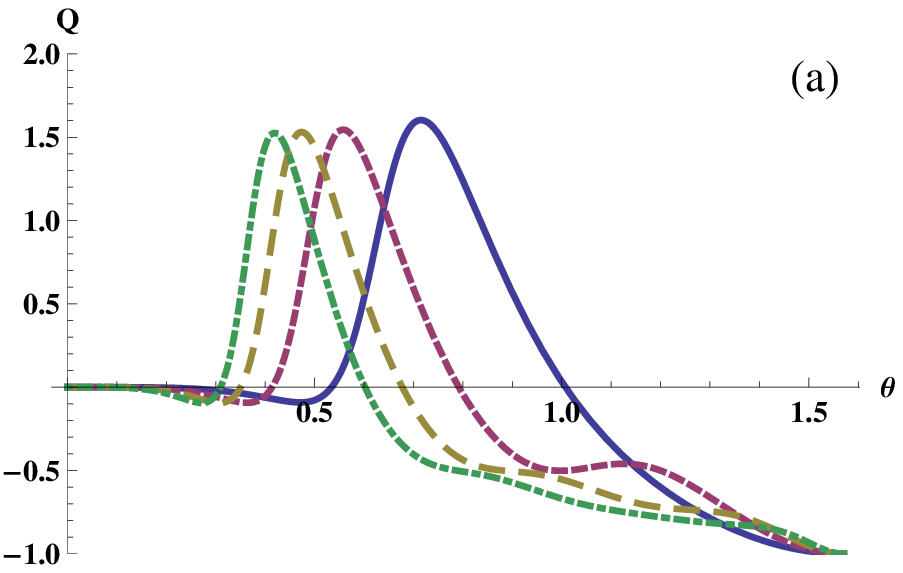} \newline%
\includegraphics[width=6cm]{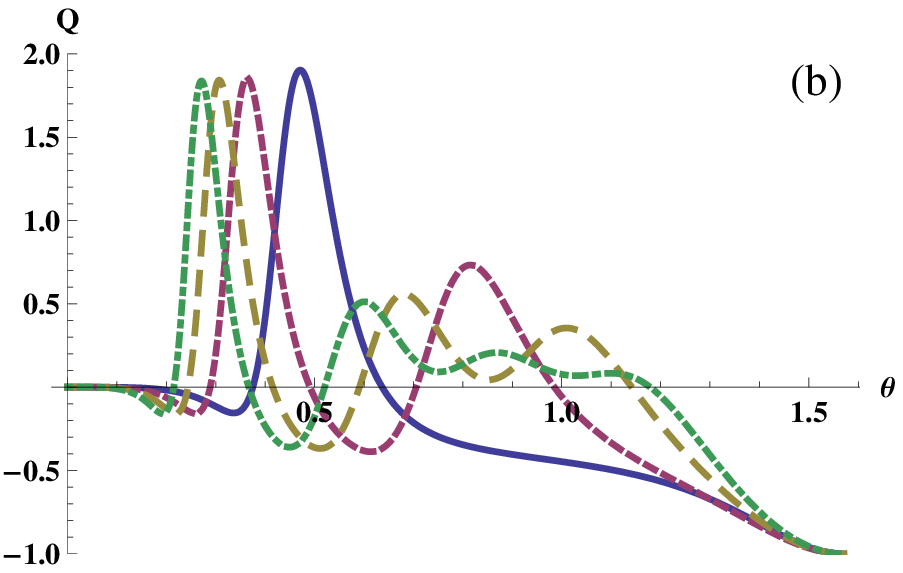}\caption{{\protect \small (Color online)
The }$\mathfrak{Q}${\protect \small -parameter as the function of
}${\protect \small \theta}${\protect \small \ for several different values
}$m=1,2,3,4,$ {\protect \small and (a) }${\protect \small z}${\protect \small =1,
(b) }${\protect \small z}${\protect \small =2. The highest peaks from right to
left correspond to m=1,2,3,4, respectively.}}%
\end{figure}

In order to see clearly the variation of Mandel's $\mathfrak{Q}$ parameter
with the input amplitude $z$ and the asymmetrical BS ($\theta$), we plot the
$\mathfrak{Q}$ parameter\ in Fig. 2 as the function of $\theta$ for some
several different values of $z$ and $m$. Here, for simplicity, we take $z$ as
a real number. From Fig. 2, we can clearly see that, for a given small $z$
value ($z=1$), the $\mathfrak{Q}$ parameter can be negative ($m\neq0$) when
$\theta$ is less than a certain threshold value or when $\theta$ is larger
than a one. Both threshold values decrease as $m$ increases; while for a large
value $z$($z=2$), the main peaks become more narrow and the corresponding
threshold values become smaller than those for the case of small value of
input amplitude. These imply that the output state presents obvious
nonclassicality which can be modulated the transmission factor.

\subsection{Second-Order Correlation Function}

Notice that the condition $\mathfrak{Q}<0$ is actually a sufficient condition
indicating\ the nonclassical property. That is to say, when $\mathfrak{Q}>0$
the state also maybe nonclassical. Next, we will further discuss the
second-order correlation function for the LPECSs, which is typically used to
find the statistical character of intensity fluctuations. The second-order
correlation function is defined by \cite{38}
\begin{equation}
g^{(2)}=\frac{\left \langle b^{\dag2}b^{2}\right \rangle }{\left \langle b^{\dag
}b\right \rangle ^{2}}=\frac{\left \langle b^{2}b^{\dag2}-4bb^{\dag
}+2\right \rangle }{\left(  \left \langle bb^{\dag}\right \rangle -1\right)
^{2}}. \label{14}%
\end{equation}
Theoretically, using the result $\left \langle b^{q}b^{\dag p}\right \rangle $
in Eq.(\ref{12}) we can get the analytical expression of $g^{(2)}$.

In Fig. 3, we plot the $g^{(2)}$ correlation function as the function of
$\theta$ for some several different values of $\left \vert z\right \vert ^{2}$
and $m$. From Fig. 3 we can see that there are some regions which present
clearly the antibunching effect with $g^{(2)}<1$, bunching effect with
$1<g^{(2)}\leqslant2$ and super-bunching effect with $2<g^{(2)}$ \cite{39}.
The antibunching effect, a nonclassical indicator, can be observed for both
high and low reflectivities (see Fig. 3 (a)). The main peaks become more
narrow and the maximal values\ of peaks become smaller than those for the
small amplitude case. The latter is different from the case of $\mathfrak{Q}%
$-parameter. In addition, the positions of peaks move to the left as the
increasing $m$. For instance, for $\left \vert z\right \vert ^{2}=1$, the peaks
corresponding to $m=1,2,3,4$ are centered around $\theta=0.68,0.53,0.45,0.39$,
which attain the corresponding measured values of $g^{(2)}%
=6.93,6.83,6.76,6.72$, respectively. It is clear that all these values of
peaks are over the limit of thermal states which is not a signature of
nonclassicality; For $m=1$, in the regions of $\theta<0.47$ and $\theta>0.90$,
the signature of nonclassicality appears and becomes more clear in the region
of $\theta>0.90$ with the increasing $\theta$. These cases are similar for
$m=2,3,4$.

\begin{figure}[ptb]
\label{Fig3} \centering \includegraphics[width=6cm]{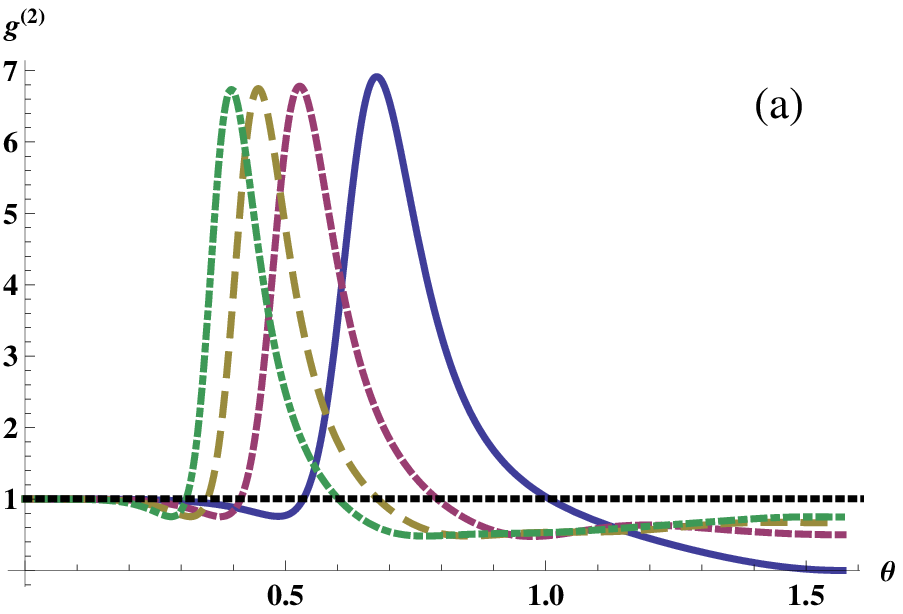} \newline%
\includegraphics[width=6cm]{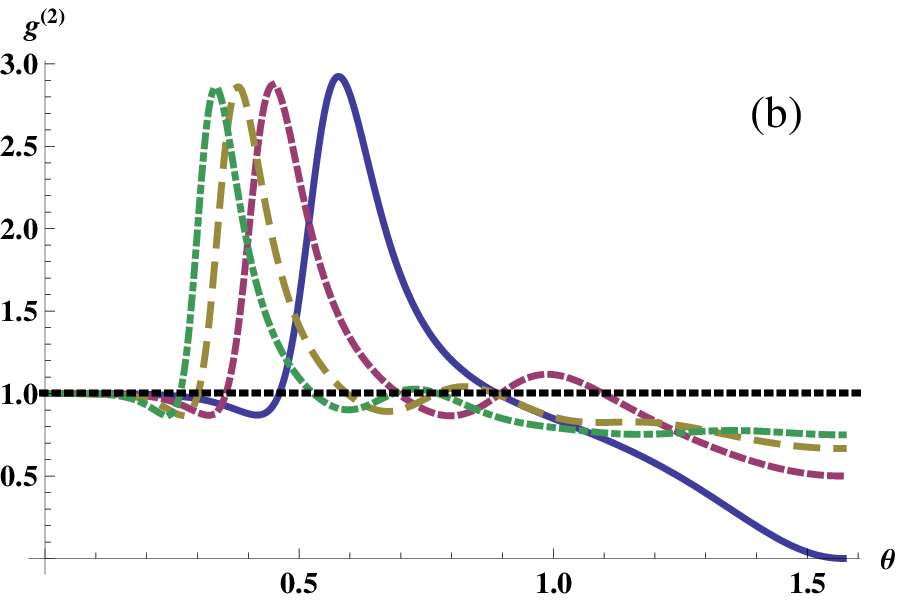}\caption{{\protect \small (Color online)
The second order correlation function }$g^{(2)}${\protect \small as the
function of }${\protect \small \theta}${\protect \small \ for several different
values }$m=1,2,3,4,$ {\protect \small and (a) }$\left \vert {\protect \small z}%
\right \vert ^{2}${\protect \small =1, (b) }$\left \vert {\protect \small z}%
\right \vert ^{2}${\protect \small =2. The highest peaks from right to left
correspond to m=1,2,3,4, respectively. When }$m=0,$$g^{(2)}=1$
{\protect \small corresponding to the classical bound.}}%
\end{figure}

\subsection{Photon number distribution}

Next, let us consider the photon-number distribution (PND) of the LPECSs. In
this field, the PND of finding $n$ photons can be calculated as%

\begin{equation}
p_{n}=\left \vert \bar{N}_{m}\right \vert ^{2}\left \vert \left \langle
n\right \vert L_{m}\left(  \mu b^{\dag}\right)  \left \vert z\cos \theta
\right \rangle \right \vert ^{2}. \label{15}%
\end{equation}
In order to obtain the explicit form of $p_{n}$, we first evaluate the matrix
element $\left \langle n\right \vert L_{m}\left(  \mu b^{\dag}\right)
\left \vert z\cos \theta \right \rangle $. Using the coherent representation of
number state in Eq.(\ref{3}) and the sum representation of Laguerre
polynomials in Eq. (\ref{9}), we have%
\begin{align}
&  \left \langle n\right \vert L_{m}\left(  \mu b^{\dag}\right)  \left \vert
z\cos \theta \right \rangle \nonumber \\
&  =\frac{1}{\sqrt{n!}}\frac{\partial^{n}}{\partial \alpha^{\ast n}}%
L_{m}\left(  \mu \alpha^{\ast}\right)  \left \langle \alpha \right \Vert
\left \vert z\cos \theta \right \rangle _{\alpha^{\ast}=0}\nonumber \\
&  =\frac{e^{-\frac{1}{2}\left \vert z\right \vert ^{2}\cos^{2}\theta}}%
{\sqrt{n!}}\frac{\partial^{n}}{\partial \alpha^{\ast n}}L_{m}\left(  \mu
\alpha^{\ast}\right)  \left.  e^{\alpha^{\ast}z\cos \theta}\right \vert
_{\alpha^{\ast}=0}\nonumber \\
&  =\frac{e^{-\frac{1}{2}\left \vert z\right \vert ^{2}\cos^{2}\theta}}%
{\sqrt{n!}}\sum_{l=0}^{m}\binom{m}{l}\frac{(-\mu)^{l}}{l!}\frac{\partial^{n}%
}{\partial \alpha^{\ast n}}\alpha^{\ast l}\left.  e^{\alpha^{\ast}z\cos \theta
}\right \vert _{\alpha^{\ast}=0}\nonumber \\
&  =\frac{e^{-\frac{1}{2}\left \vert z\right \vert ^{2}\cos^{2}\theta}}%
{\sqrt{n!}}\sum_{l=0}^{m}\binom{m}{l}\binom{n}{l}(-\mu)^{l}\left(  z\cos
\theta \right)  ^{n-l}, \label{16}%
\end{align}
thus the PND is given by%
\begin{align}
p_{n}  &  =\frac{\left \vert \bar{N}_{m}\right \vert ^{2}}{n!}e^{-\left \vert
z\right \vert ^{2}\cos^{2}\theta}\nonumber \\
&  \times \left \vert \sum_{l=0}^{m}\binom{m}{l}\binom{n}{l}(-\mu)^{l}\left(
z\cos \theta \right)  ^{n-l}\right \vert ^{2}. \label{17}%
\end{align}
It is easy to see that Eq.(\ref{17}) just reduces to the PND of the coherent
state $\left \vert z\cos \theta \right \rangle $ when $m=0$, i.e., $p_{n}=\frac
{1}{n!}e^{-\left \vert z\right \vert ^{2}\cos^{2}\theta}\left \vert z\cos
\theta \right \vert ^{2n}$, as expected.

\begin{figure}[ptb]
\label{Fig4} \centering \includegraphics[width=4.2cm]{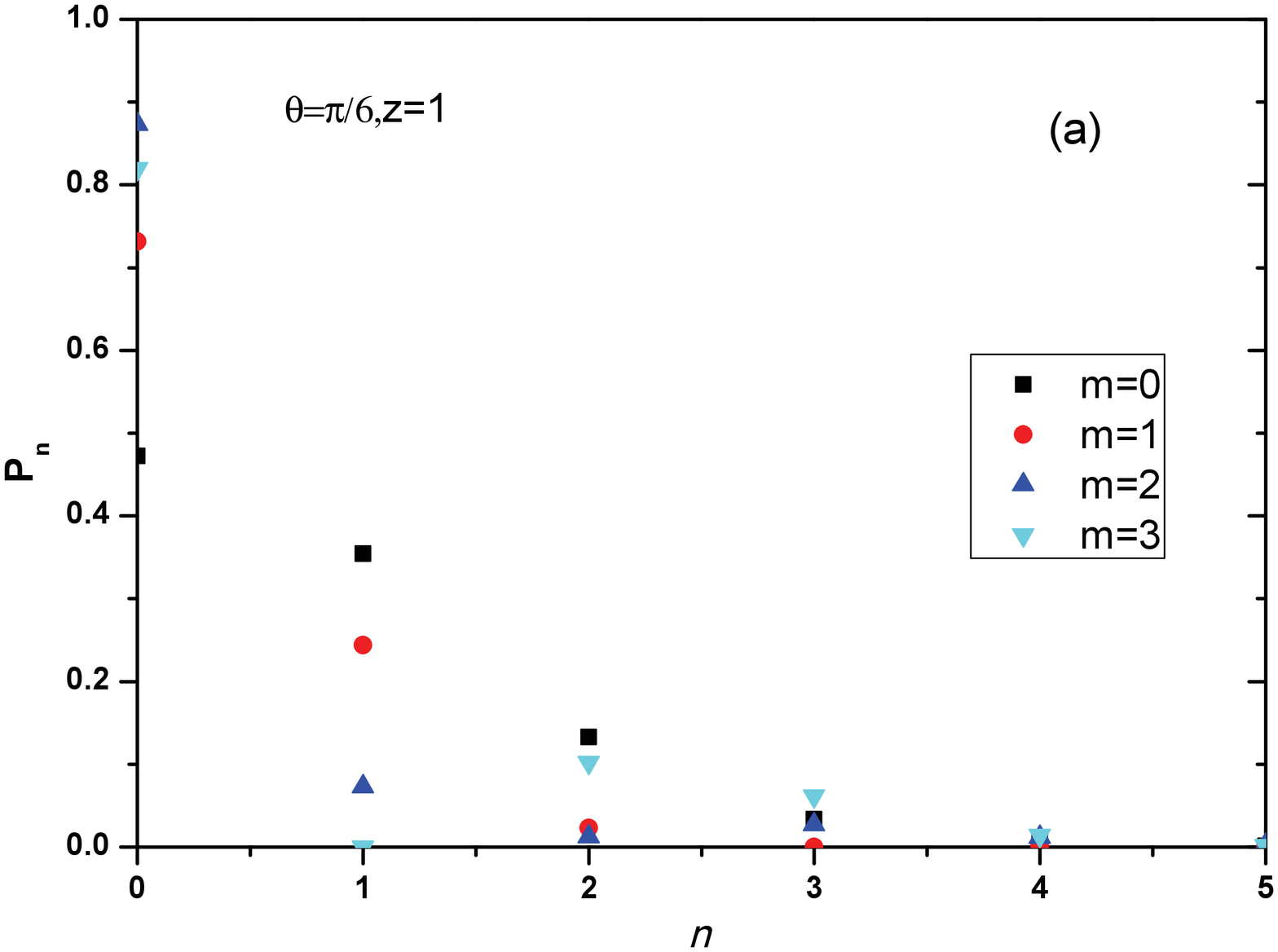}
\includegraphics[width=4.2cm]{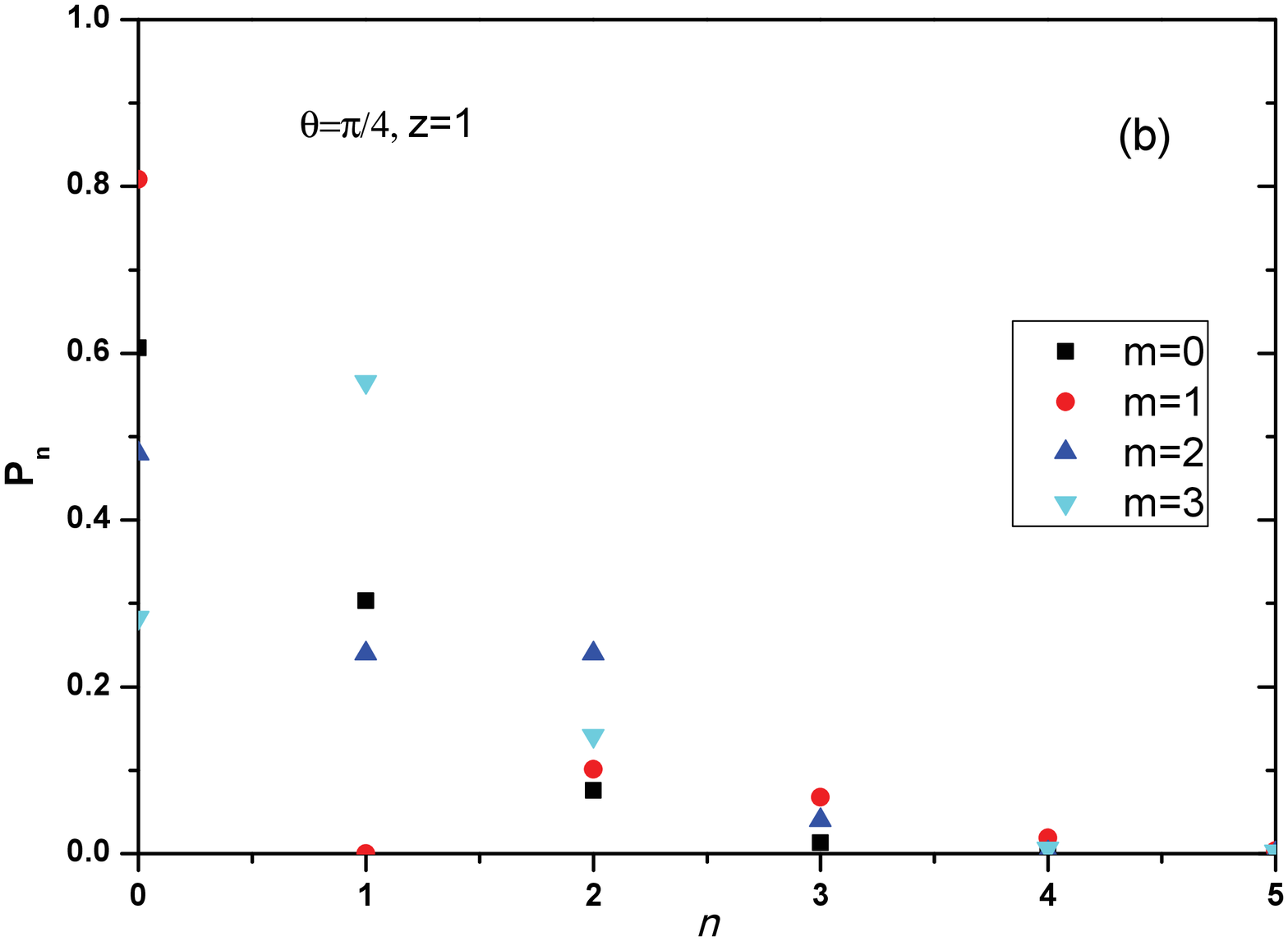} \  \  \newline%
\includegraphics[width=4.2cm]{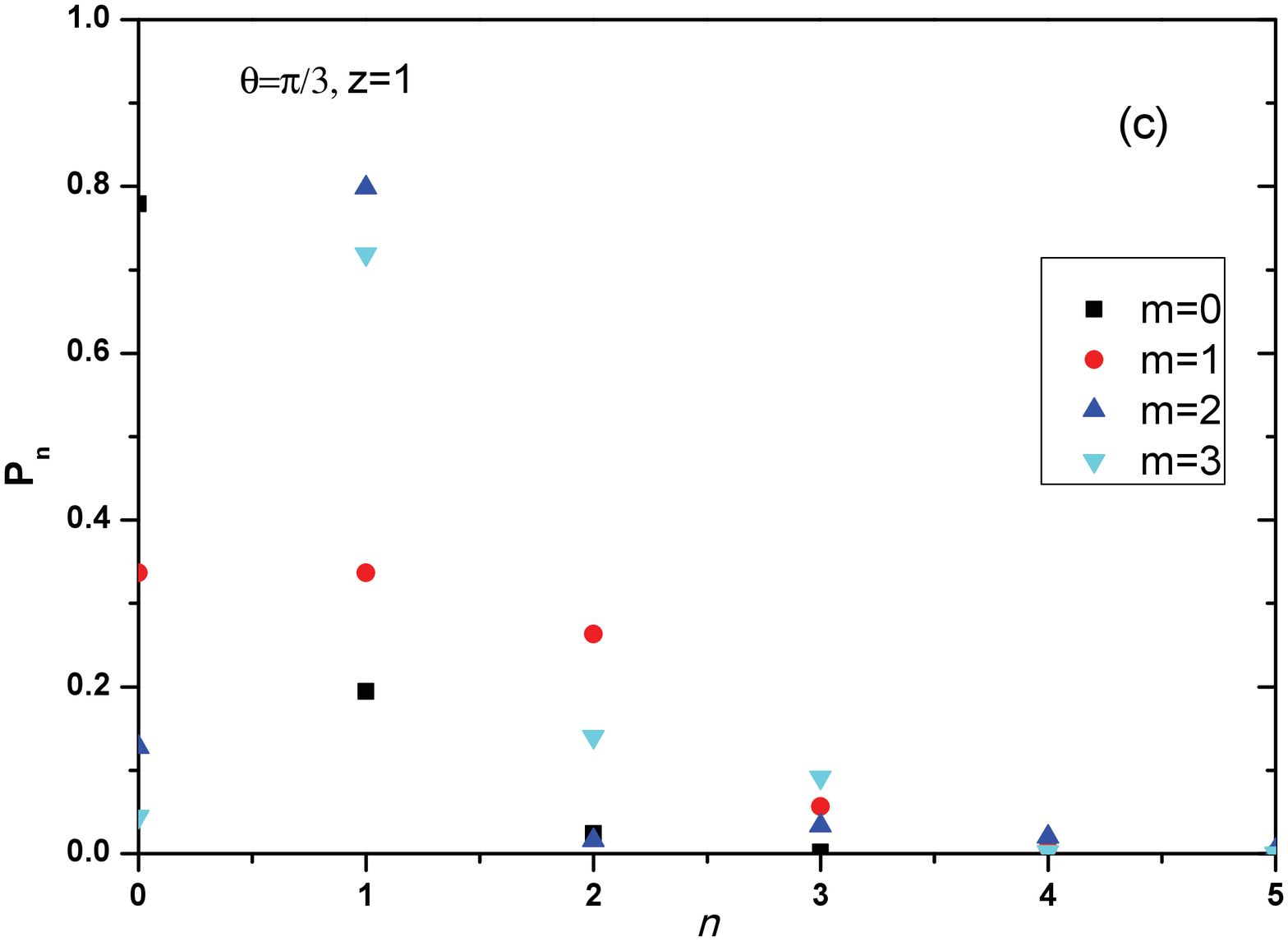}
\includegraphics[width=4.2cm]{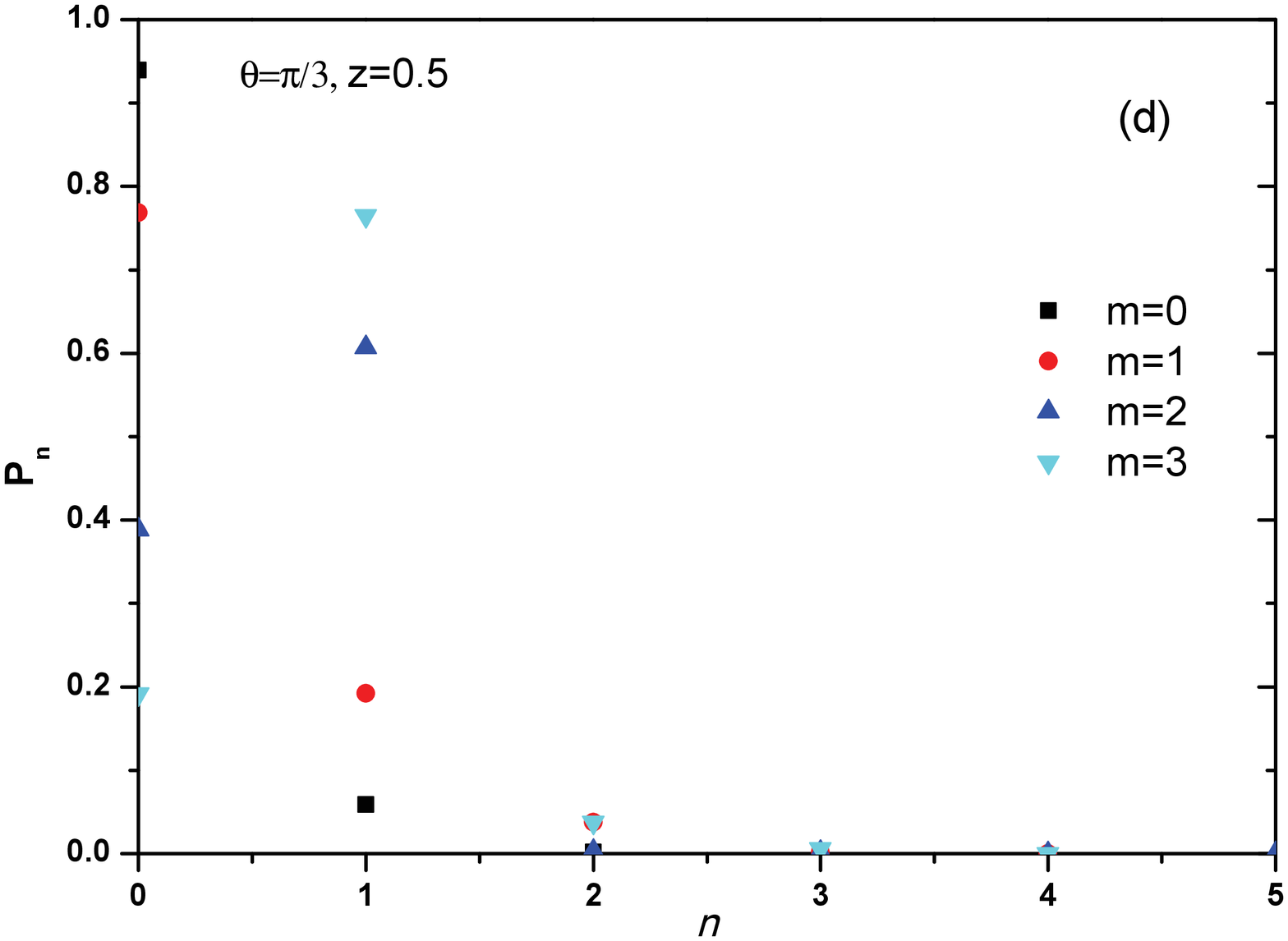}\caption{{}(Color online) The
photon-number distribution of the LPECSs as a function of $n$ for several
different parameters $\theta$, $m$ and $z$. (a) $\theta=\pi/6,z=1;$ (b)
$\theta=\pi/4,z=1;$ (c) $\theta=\pi/3,z=1;$ (d) $\theta=\pi/3,z=0.5.$}%
\end{figure}

In Fig. 4, the PND is plotted for several different parameters $\theta$, $m$
and $z$, from which we can see that (i) the peak of PND is mainly located at
$n=0$ for the case of $m=0$ and different values of $\theta$ and $z$ (see
Figs. 4.(a)-(d)); (ii) by modulating the order $m$ of Laguerre polynomials, we
may change the position and value of the peak. For example, the maximum values
of peaks at $n=0$ increase as $m$ increase (see Fig. 4(a)); (iii) for
$m=1,2,3$, the PND is mainly distributed at $n=1$ and the maximum values of
peaks modulated by beam splitter ($\theta$), which implies that we can prepare
single photon Fock state by this conditional measurement for a given amplitude
of input coherent state; for instance, when $\theta=\pi/4$ and $m=3$, we can
get a single-photon in a success probability of $0.57$ (see Fig. 4(b)), while
for $\theta=\pi/3$ the probabilities are 0.80 and 0.72 for $m=2,3$ (see Fig.
4(c)), respectively. This is to say, we can achieve the single-photon at a
smaller measured $m$ when increasing the value of $\theta$ for a given $z$;
(iv) for a small amplitude value of z (see Fig. 4(d)), we can increase the
measured $m$ to obtain the single-photon in a higher probability (say when
$m=2,3$, probability=0.61 and 0.77). Thus we can not only modify the PND but
also achieve single-photon Fock state by the quantum catalysis rather than
photon-subtraction (photon-loss).

\subsection{Squeezing properties}

Now, we investigate the squeezing properties of the LPECSs via the quadrature
variance $\left(  \triangle Q\right)  ^{2}<1$ or $\left(  \triangle P\right)
^{2}<1$ which indicates the squeezing or sub-Poissonian statistics. The
quadrature components of the optical field is given by $Q=(b+b^{\dag}%
)/\sqrt{2}$ and $P=(b-b^{\dag})/(i\sqrt{2})$. Thus the quadrature variances
can be expressed as the following anti-normally ordering forms
\begin{align}
\left(  \triangle Q\right)  ^{2}  &  =\left \langle Q^{2}\right \rangle
-\left \langle Q\right \rangle ^{2}\nonumber \\
&  =\frac{1}{2}\left \{  \left \langle b^{2}\right \rangle -\left \langle
b\right \rangle ^{2}+\left \langle b^{\dag2}\right \rangle -\left \langle b^{\dag
}\right \rangle ^{2}\right. \nonumber \\
&  \left.  +2\left \langle bb^{\dag}\right \rangle -2\left \langle b\right \rangle
\left \langle b^{\dag}\right \rangle -1\right \}  , \label{18}%
\end{align}
and%
\begin{align}
\left(  \triangle P\right)  ^{2}  &  =\left \langle P^{2}\right \rangle
-\left \langle P\right \rangle ^{2}\nonumber \\
&  =\frac{1}{2}\left \{  -\left \langle b^{2}\right \rangle +\left \langle
b\right \rangle ^{2}-\left \langle b^{\dag2}\right \rangle +\left \langle b^{\dag
}\right \rangle ^{2}\right. \nonumber \\
&  \left.  +2\left \langle bb^{\dag}\right \rangle -2\left \langle b\right \rangle
\left \langle b^{\dag}\right \rangle -1\right \}  . \label{19}%
\end{align}
Using Eq.(\ref{12}) we can get the analytical expressions for the variance of
the quadratures, but are not given here. Next, we shall discuss the squeezing
properties by\ numerical calculation.

In Fig. 5, we plot these optimal quadrature variances as a function of the
input amplitudes for several different values of $m$ by minimizing variances
$\left(  \triangle Q\right)  ^{2}$ over $\theta$ from 0 to $\pi/2$. Here, we
take a logarithmic scale, i.e. units of dB whose definition is given by
dB$[Q]=10\log_{10}[\left(  \triangle Q\right)  ^{2}/\left(  \triangle
Q\right)  _{vac}^{2}]$ and dB$[P]=10\log_{10}[\left(  \triangle P\right)
^{2}/\left(  \triangle P\right)  _{vac}^{2}],$ where $\left(  \triangle
Q\right)  _{vac}^{2}$ and $\left(  \triangle P\right)  _{vac}^{2}$
corresponding to the vacuum variances of $1/2$ for our definition of
quadrature components. In Fig. 5(a), it is clearly seen that (i) when $m=1$,
the optimal squeezing is 1.249 dB below the shot-noise limit and it is
independent of the input amplitude $z$; (ii) the optimal values of squeezing
or the minimum variances monotonously increase as $m$ for a given amplitude
$z$, and decrease as $z$ for $m=2,3,4$. These results indicate that the
squeezing can be enhanced by increasing measured $m$ photons or reducing the
amplitude $z$. Fig. 5(b) shows the $\theta$ values corresponding to the
largest squeezing effect as a function of $z$, from which we can see that the
$\theta$ value decreases as $m$ for a given $z$ and monotonously decreases as
$z$ for a given $m$.

\begin{figure}[ptb]
\label{Fig5} \centering \includegraphics[width=7cm]{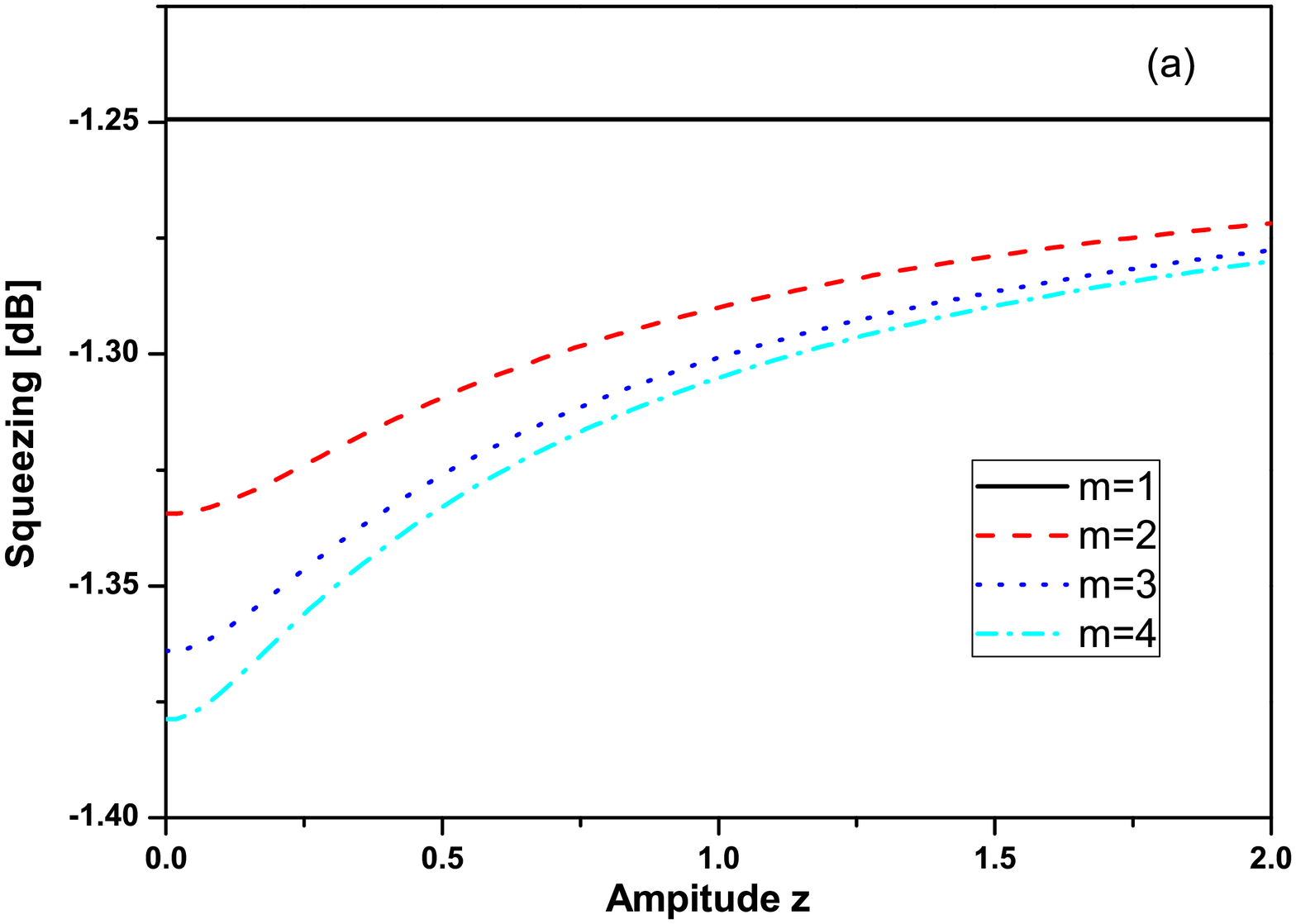} \newline%
\includegraphics[width=7cm]{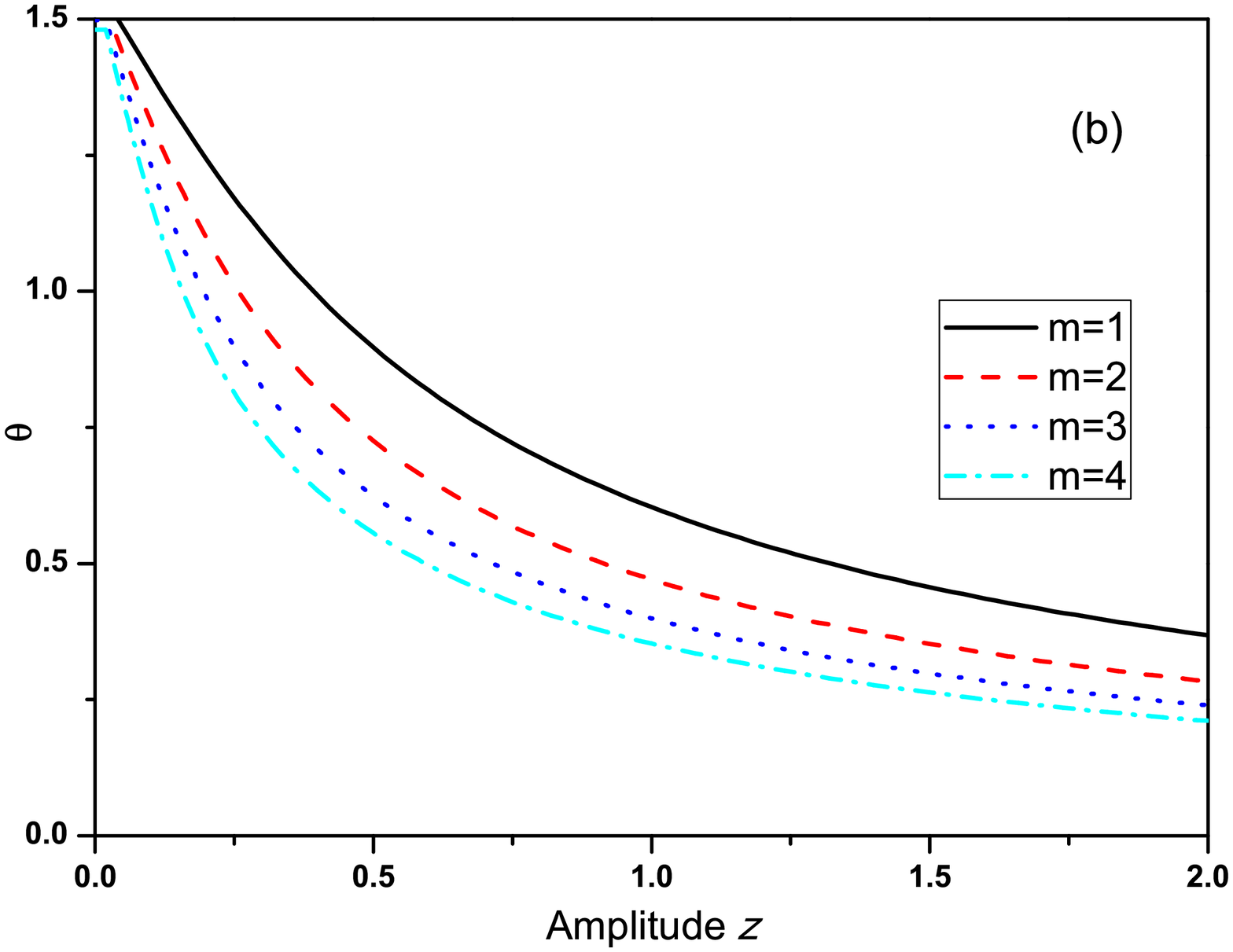} \newline \caption{{\protect \small (Color
online) (a) The optimal squeezing effect as a function of the input amplitude
}$z$ {\protect \small by minimizing the variances of quadrature component }%
$Q${\protect \small over }$\theta${\protect \small ; (b) the }$\theta
${\protect \small values corresponding to the optimal squeezing effect as a
function of }$z${\protect \small .}}%
\end{figure}

Next, we further consider the squeezing properties of the LPECSs by
introducing another quadrature operator $Q_{\varphi}=ae^{-i\varphi}%
+a^{\dagger}e^{i\varphi}$. Thus the squeezing can be characterized by the
minimum value $\left \langle \triangle^{2}Q_{\varphi}\right \rangle $ $<1$ with
respect to $\varphi$, or by the normal ordering form $\left \langle
\colon \triangle^{2}Q_{\varphi}\colon \right \rangle <0$. Upon expanding the
terms of $\left \langle \colon \triangle^{2}Q_{\varphi}\colon \right \rangle $,
one can minimize its value over the whole angle $\varphi$. The optimized
nonclassical depth over the phases is found to be \cite{40}%
\begin{equation}
S_{opt}=-2\left \vert \left \langle a^{\dagger2}\right \rangle -\left \langle
a^{\dagger}\right \rangle ^{2}\right \vert +2\left \langle aa^{\dagger
}\right \rangle -2\left \vert \left \langle a^{\dagger}\right \rangle \right \vert
^{2}-2. \label{20}%
\end{equation}
The negative value of $S_{opt}$ in the range $\left[  -1,\text{ }0\right)  $
implies squeezing (or nonclassical). Using Eq.(\ref{12}) we can get the
expression of $S_{opt}$. In particular, when $m=0$ (the case of coherent
output), $S_{opt}=0$, as expected. In Fig. 6 we plot the $S_{opt}$ as a
function of $\theta$ for some different values of $m$ and $z$, from which we
can see that there is a region of $\theta$ for representing the negative value
of $S_{opt},$ and the region becomes smaller with the increasing $m$ and $z.$
For a given $m$ or $z$, the region becomes narrower for a bigger $z$ or $m$.
For more discussions about higher-order nonclassical effects of quantum state,
we refer to Refs.\cite{41,42,43,44}.

\begin{figure}[ptb]
\label{Fig6}
\centering \includegraphics[width=6cm]{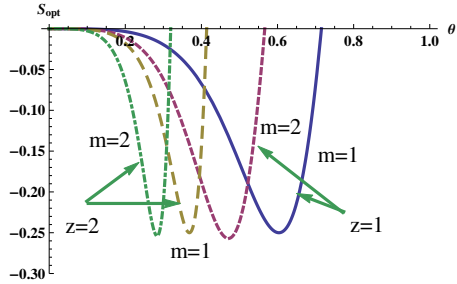}\caption{{\protect \small (Color
online) The degree of squeezing Sopt of the LPECSs as the function of }%
$\theta$ {\protect \small for different values of }$m${\protect \small and }%
$z${\protect \small .}}%
\end{figure}

\section{Wigner distribution of the LPECSs}

In this section, we shall discuss the quasi-probability distribution, Wigner
function, whose negativity may be considered as a good indicator of the
nonclassicality. For the single-mode case, the Wigner function can be
calculated as
\begin{equation}
W\left(  \gamma \right)  =\mathtt{tr}(\rho \Delta(\gamma)), \label{21}%
\end{equation}
where $\Delta(\gamma)$ is the single-mode Wigner operator \cite{45}, defined
by%
\begin{equation}
\Delta(\gamma)=e^{2\left \vert \gamma \right \vert ^{2}}\int \frac{d^{2}\alpha
}{\pi^{2}}\left \vert \alpha \right \rangle \left \langle -\alpha \right \vert
e^{-2(\alpha \gamma^{\ast}-\gamma \alpha^{\ast})}, \label{22}%
\end{equation}
and $\left \vert z\right \rangle =\exp \left(  zb^{\dag}-z^{\ast}b\right)
\left \vert 0\right \rangle $ is Glauber coherent state. Substituting
Eqs.(\ref{5}) and (\ref{22}) into Eq.(\ref{21}) and using Eq.(\ref{7}), the
Wigner function of can be derived as%
\begin{equation}
W_{m}\left(  \gamma \right)  =\left \vert \bar{N}_{m}\right \vert ^{2}%
e^{2\left \vert \gamma \right \vert ^{2}-\left \vert z\right \vert ^{2}\cos
^{2}\theta}\Theta \left(  \mu,\mu^{\ast}\right)  , \label{23}%
\end{equation}
where
\begin{align}
\Theta \left(  \mu,\mu^{\ast}\right)   &  =\int \frac{d^{2}\alpha}{\pi^{2}}%
L_{m}\left(  -\mu \alpha^{\ast}\right)  L_{m}\left(  \mu^{\ast}\alpha \right)
\nonumber \\
&  \times e^{-\left \vert \alpha \right \vert ^{2}+\alpha \left(  z^{\ast}%
\cos \theta-2\gamma^{\ast}\right)  -\alpha^{\ast}\left(  z\cos \theta
-2\gamma \right)  }. \label{24}%
\end{align}
It is easy to see that the Wigner function $W\left(  \gamma \right)  $ is a
real number in phase space, since $\Theta^{\ast}\left(  \mu,\mu^{\ast}\right)
=\Theta \left(  \mu,\mu^{\ast}\right)  $.

Furthermore, using Eqs.(\ref{9}) and (\ref{11}) we can finally obtain the
Wigner function%
\begin{equation}
W_{m}\left(  \gamma \right)  =W_{0}\left(  \gamma \right)  F_{m}\left(
\gamma \right)  , \label{25}%
\end{equation}
where $W_{0}\left(  \gamma \right)  =1/\pi \exp \{-2\left \vert \gamma-z\cos
\theta \right \vert ^{2}\}$ is just the Wigner function of coherent state
$\left \vert z\cos \theta \right \rangle $, and the non-Gaussian item
$F_{m}\left(  \gamma \right)  $ is defined by%
\begin{align}
F_{m}\left(  \gamma \right)   &  =\left \vert \bar{N}_{m}\right \vert ^{2}%
\sum_{j,l=0}^{m}\frac{\mu^{l}\mu^{\ast j}}{l!j!}\binom{m}{j}\binom{m}%
{l}\nonumber \\
&  \times H_{l,j}\left(  z^{\ast}\cos \theta-2\gamma^{\ast},z\cos \theta
-2\gamma \right)  , \label{26}%
\end{align}
which is from the presence of conditionally measured $m$ photons. In
particular, when $m=1,$ then we have
\begin{align}
F_{1}\left(  \gamma \right)   &  =\left \vert \bar{N}_{1}\right \vert
^{2}\{1+\left \vert \mu \right \vert ^{2}(1-\left \vert z\cos \theta-2\gamma
\right \vert ^{2})\nonumber \\
&  +[\mu \left(  z^{\ast}\cos \theta-2\gamma^{\ast}\right)  +c.c]\}. \label{27}%
\end{align}
The negative region of Wigner function will be decided by $F_{1}\left(
\gamma \right)  <0$. \begin{figure}[ptb]
\label{Fig7}
\centering \includegraphics[width=9cm]{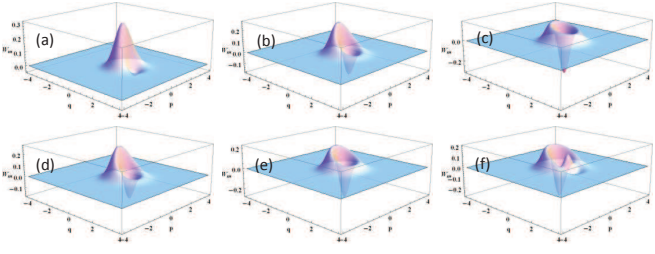}\caption{{\protect \small (Color
online) The Wigner function distribution in phase space with z=1 for some
different values of m and }$\theta${\protect \small . The first and second rows
correspond to m=1,2, respectively. And }$\theta \ ${\protect \small are equal to
}$\pi/5,\pi/4,\pi/3${\protect \small from left to right for each column.}}%
\end{figure}

In Fig. 7, we plot the Wigner distributions in phase space for several
different parameter values of $m$ and $\theta$ with $z=1$, from which it is
clearly seen that there are some obvious negative regions of the Wigner
function in the phase space which is an indicator of the nonclassicality of
the state. Furthermore, these negative areas are modulated not only by $m$,
but also by $\theta$. For example, for a given $\theta=\pi/5$ (see Figs. 7 (a)
and (d)), there is a bigger negative volume of the Wigner function for the
case of $m=2$ than that for $m=1$; and for a given $m=1$, the negative volume
of the Wigner function becomes bigger as $\theta$ increases (see Fig.7
(a)-(c)). Actually, the structure of the Wigner function will be affected by
the input amplitude $z$. In order to clearly see these above points, we
further quantify the negative volume of the Wigner function, defined by
$\delta=\frac{1}{2}[\int_{-\infty}^{\infty}dqdp\left \vert W(q,p)\right \vert
-1]$ with $\gamma=(q+ip)/\sqrt{2}$. In table I, we present some values of
negative volume of the Wigner function for different $m$, $\theta$, and $z$.
It is clearly seen that the effects on the nonclassicality are different due
to the changing of parameters $m$, $\theta$, and $z$.%

\begin{tabular}
[c]{c||ccc||ccc}%
\multicolumn{7}{c}{}\\
\multicolumn{7}{c}{TABLE I: Negative volume $\delta$\ of the WF}\\ \hline \hline
case & \multicolumn{3}{c||}{z=1} & \multicolumn{3}{|c}{z=2}\\ \hline \hline
case & $\theta$=$\frac{\pi}{5}$ & $\theta$=$\frac{\pi}{4}$ & $\theta$%
=$\frac{\pi}{3}$ & $\theta$=$\frac{\pi}{5}$ & $\theta$=$\frac{\pi}{4}$ &
$\theta$=$\frac{\pi}{3}$\\ \hline \hline
m=1 & 0.023 & 0.115 & 0.205 & 0.163 & 0.115 & 0.122\\ \hline
m=2 & 0.116 & 0.180 & 0.207 & 0.149 & 0.271 & 0.297\\ \hline
m=3 & 0.164 & 0.188 & 0.212 & 0.242 & 0.307 & 0.412\\ \hline \hline
\end{tabular}

\bigskip

\section{Decoherence in thermal environment}

In this section, we consider the decoherence of the LPECSs by analytically
deriving the Wigner function in thermal environment. When the quantum state
evolves in a thermal environment associated with Born-Markivian approximation,
the evolution of density operator can be described by the following master
equation \cite{46}
\begin{align}
\frac{d\rho}{dt}  &  =\kappa(\tilde{n}+1)(2a\rho a^{\dagger}-a^{\dagger}%
a\rho-\rho a^{\dagger}a)\label{28}\\
&  +\kappa \tilde{n}\left(  2a^{\dagger}\rho a-aa^{\dagger}\rho-\rho
aa^{\dagger}\right)  ,\nonumber
\end{align}
where $\kappa$ denotes the dissipative coefficient and $\tilde{n}$ represents
the average thermal photon number of the lossy channel. Using the entangled
state representation, we have derived the sum representation of Klaus operator
and the evolution of the Wigner function governed by Eq.(\ref{28})
\cite{47,48}. The latter is given by%
\begin{equation}
W\left(  \beta,\beta^{\ast},t\right)  =\frac{2}{(2\tilde{n}+1)T}\int
\frac{d^{2}\gamma}{\pi}W\left(  \gamma,\gamma^{\ast}\right)  e^{-2\frac
{|\beta-\gamma e^{-\kappa t}|^{2}}{(2\tilde{n}+1)T}}, \label{29}%
\end{equation}
where $W\left(  \gamma,\gamma^{\ast}\right)  $ is the initial Wigner function
and $T=1-e^{-2\kappa t}$. Then substituting Eq.(\ref{23}) into Eq.(\ref{29})
and using the following integration formula
\begin{equation}
\int \frac{d^{2}z}{\pi}e^{-\zeta \left \vert z\right \vert ^{2}+\xi z+\eta
z^{\ast}}=\frac{1}{\zeta}e^{\frac{\xi \eta}{\zeta}},\text{Re}\left(
\zeta \right)  >0, \label{29a}%
\end{equation}
we finally obtain%
\begin{equation}
W\left(  \beta,\beta^{\ast},t\right)  =W_{0}\left(  \beta,t\right)
F_{m}\left(  \beta,t\right)  , \label{30}%
\end{equation}
where $W_{0}\left(  \beta,t\right)  =\frac{1}{\pi A}\exp \{-\allowbreak
2\left \vert \beta-\bar{z}e^{-\kappa t}\right \vert ^{2}/A\}$ with $A=2\tilde
{n}T+1\ $and $\bar{z}=z\cos \theta$ is the evolution of Wigner function of
coherent state $\left \vert z\cos \theta \right \rangle $ in thermal channel, and
\begin{align}
F_{m}\left(  \beta,t\right)   &  =\left \vert \bar{N}_{m}\right \vert ^{2}%
\sum_{j,l=0}^{m}\frac{\mu^{l}\mu^{\ast j}}{l!j!}\binom{m}{j}\binom{m}%
{l}\left(  \sqrt{\frac{B}{A}}\right)  ^{l+j}\nonumber \\
&  \times H_{l,j}\left(  \frac{\bar{z}^{\ast}B-2\beta^{\ast}e^{-\kappa t}%
}{\sqrt{AB}},\frac{\bar{z}B-2\beta e^{-\kappa t}}{\sqrt{AB}}\right)
\nonumber \\
&  \left(  B=e^{-2\kappa t}-(2\tilde{n}+1)T\right)  . \label{31}%
\end{align}
In particular, when $t=0$ Eqs.(\ref{30}) and (\ref{31}) just reduce to
Eqs.(\ref{25}) and (\ref{26}), respectively.

In Fig. 8, we take the case with $m=1$ and $z=1$ as well as $\theta=\pi/3$ as
an example for the evolution of Wigner function. From Fig. 8, due to the
presence of decoherence, the negative region of Wigner function gradually
disappears with the incasement of $\kappa t$. In addition, the characteristic
time $\kappa t_{c}$, which means that there is always negative region for
Wigner function in phase space when the decay time is less than $\kappa t_{c}%
$, is dependent not only on the catalytic photon number $m$, but also on the
reflectivity of unbalanced BS. In order to see clearly this point, we plot the
minimum negative values of Wigner function as a function of decay time $\kappa
t$ in Fig. 9, from which it is clear that the minimum negative decreases
monotonously as $\kappa t$; in addition, for instance, the characteristics
times for $m=1$ are about $0.20,0.27$ and $0.30$ corresponding to $\theta
=\pi/5,\pi/4,\pi/3$, respectively. There is a longer characteristic time for
$\theta=\pi/3$ than that for $\theta=\pi/5,\pi/4.$ This point can be
understood like that: because the case of $\theta=\pi/3$ has a higher
reflectivity than that of $\theta=\pi/5,\pi/4$, thus the output state presents
more properties of number state. Actually, the output state can be considered
as the superposition of coherent state and number state in a certain form.

\begin{figure}[ptb]
\label{Fig8} \centering \includegraphics[width=8cm]{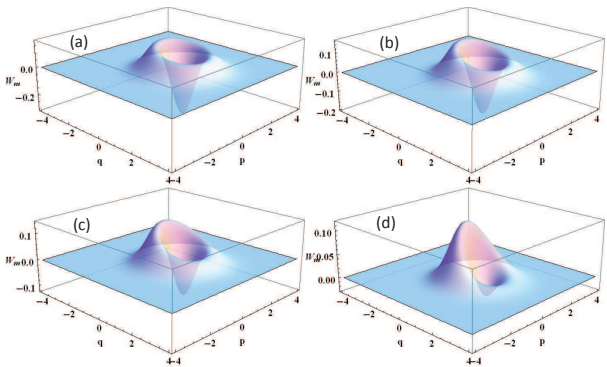}\caption{{}(Color
online) The evolution of Wigner function distribution in phase space with
$m=1,$ $\tilde{n}=0$. (a)-(d) $\kappa t=0,0.05,0.1,0.2.$}%
\end{figure}

\begin{figure}[ptb]
\label{Fig9} \centering \includegraphics[width=8cm]{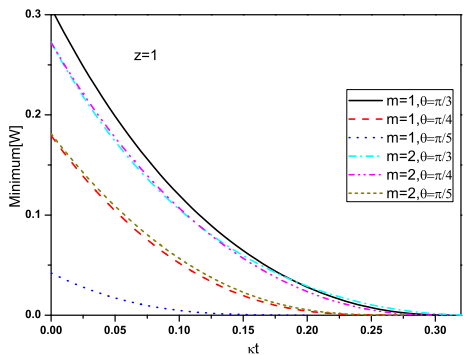}\caption{{}(Color
online) The minimum value of negative Wigner function distribution in phase
space as the function of the decay time $\kappa t$ for $m=1,2,$ $\tilde{n}=0$
and $z=1,$ $\theta=\pi/5,\pi/4,\pi/3.$}%
\end{figure}

\section{Conclusion}

In this paper, we proposed a new kind of non-Gaussian state----Laguerre
polynomial excited coherent state by using multiphoton catalysis which is
proposed firstly by Lvovsky and Mlynek. It is shown that the multiphoton
catalysis can actually be seen as a block comprising photon number operator.
We then considered the nonclassical properties of the LPECs when considering
the coherent state as inputs. It is found that the state can present
sub-Possion statistics, antibunching effect and squeezing behavior. All these
properties can be modulated by the amplitude of coherent state, catalysis
number and unbalanced BS. In particular, the maximum squeezing for the case of
$m=1$ is kept to be constant ($\sim$1.249dB) by optimizing over unbalanced BS.
The maximum squeezing can be improved by increasing $m$ and reducing the
amplitude of coherent state. In addition, we also examined the decoherence
behavior of the LPECs according to the negativity of Wigner function. It is
found that the negative region, characteristic time of decoherence and
structure of Wigner function are affected by catalysis number and unbalanced BS.

Here the generation of Laguerre polynomials excited state is just an example
for opening the way approaching a series of non-Gaussian quantum state.
Actually, by different herald inputs and different measurements, we can
achieve some other non-Gaussian states such as Hermite polynomials excited
squeezed states, etc. Our current work provides a general analysis about how
to prepare theoretically such polynomials quantum states.

It would be interesting to extend this work to multi-mode case including how
to realize the entanglement distillation and improve the fidelity of
teleportation. On the other hand, non-Gaussian quantum states have a wide
application in quantum information and quantum computation \cite{49}. For
example, by using photon-subtraction operator, a scheme is proposed to improve
the performance of entanglement-based continuous-variable
quantum-key-distribution protocol \cite{12}. It is found that the subtraction
operation can increase the secure distance and tolerable excess noise of the
entanglement-based scheme, as well as the corresponding prepare-and-measure
scheme. Recently, for another example, the single-photon-added coherent state
has been used in quantum key distribution \cite{13}. It is shown that the
single-photon-added coherent source can greatly exceed all other existing
sources in both BB84 protocol and the recently proposed
measurement-device-independent quantum key distribution. These investigations
are good examples for showing that it is possible to enhance the performance
in the field of quantum information by preparing various non-Gaussian states.
Thus, the applications of such non-Gaussian states including the LPECSs with
continuous-variable in quantum information could be paid attention in the future.

\textbf{Acknowledgments: }This work is supported by a grant from the Qatar
National Research Fund (QNRF) under the NPRP project 7-210-1-032. L. Y. Hu is
supported by the China Scholarship Council (CSC) and the National Natural
Science Foundation of China (Nos.11264018 and 11264015), as well as the
Natural Science Foundation of Jiangxi Province (No. 20152BAB212006), the
Research Foundation of the Education Department of Jiangxi Province of China
(no GJJ14274).

\end{document}